\documentstyle[11pt,newpasp,twoside,epsf]{article}
\def\lesssim{{_ <\atop{^\sim}}}
\def\grtsim{{_ >\atop{^\sim}}}

\def\lesssim{{_ <\atop{^\sim}}}
\def\grtsim{{_ >\atop{^\sim}}}
\def\msun{\mbox{M$_\odot$}}
\def\fd{\mbox{f$_d$}}

\def\vmax{\mbox{V$_m$}}

\def\edcomment#1{\iffalse\marginpar{\raggedright\sl#1\/}\else\relax\fi}
\reversemarginpar

\begin{document}
\title{The shape of rotation curves: models vs. observations}

\author{Vladimir Avila-Reese\altaffilmark{1}, Claudio Firmani\altaffilmark{2} 
and Jes\'us Zavala\altaffilmark{1,3}}

\altaffiltext{1}{Instituto de Astronom\'\i a, UNAM, A.P. 70-264, 04510 M\'exico, D. F.}
\altaffiltext{2}{Osservatorio Astronomico di Brera, via E.Bianchi 46, 
I-23807 Merate, Italy}
\altaffiltext{3}{Facultad de Ciencias, UNAM,  04510 M\'exico, D. F.}

\begin{abstract}

We discuss the shape and decomposition of rotation curves (RCs) of galaxies
formed within growing cold dark matter halos. The outer RC shape correlates 
mainly with the surface brightness (SB), the luminous mass fraction, \fd, 
and the bulge fraction. In order the shapes of RC depend significantly on 
luminosity, \fd\ should be a strong function of mass (feedback). For the preferred 
values of \fd\ ($\lesssim 0.03$), the high SB 
models can be maximum disks only when the halos have a shallow core. 
The low SB models are sub-maximum disks. The residuals of the baryonic 
Tully-Fisher (TF) and mass-radius relations show a clear anti-correlation, but 
when one passes to the TF and L-R relations, the anti-correlation almost 
disappears. Therefore, the observed lack of correlation among the 
residuals of the last two relations should not be interpreted as an evidence
of sub-maximal disks.

\end{abstract}

\index{galaxies: formation --- galaxies: haloes --- galaxies: kinematics
and dynamics ---  galaxies: spiral --- dark matter}

\section{Introduction}

The shape and amplitude of the rotation curves (RCs) reflect the 
mass distribution of luminous and dark matter in disk galaxies.
The luminous mass distribution is inferred from the observed surface 
brightness and gas profiles after assuming a stellar mass-to-luminosity 
ratio M/L. The uncertainty in the determination of the M/L
ratio is reflected in the large number of RC decomposition models 
proposed, from maximum disk hypothesis to sub-maximum disks to minimum 
disks.
Whether luminous or dark matter dominates can in a first approximation
be inferred from the shape of the RCs. If the luminous mass contribution to 
the gravitational potential is dynamically important, then the shapes of 
the RCs are expected to differ among galaxies with different surface 
brightness (SB) distributions. The dark matter halo modulates these differences
and in the extreme case of dark halo dominion, the shapes of the RCs will be 
largely independent from the SB profile. 

In this paper we will present the shape and decomposition of RCs of modeled
disk galaxies within cold dark matter (CDM) halos and discuss these results in 
the light of the observations.

\section{The model} 

The disk is build up within a growing CDM halo. We use the extended Press-Schechter 
approach to generate the statistical mass aggregation histories (MAHs) of 
the halos, and a generalized secondary infall model to calculate the 
virialization of the accreting mass shells (Avila-Reese et al. 1998). The 
evolution and structure of our CDM halos agree well with results from
cosmological N-body simulations. The mass shells are assumed to have aligned 
rotation axis and to be in solid body  rotation, with specific angular 
momentum given by $j_{sh}(t_v)=dJ(t_v)/dM_v(t_v)$, where 
$J=\lambda GM_v^{5/2}/\left| E\right| ^{1/2}$, $J$, M$_v$ and $E$ are
the total angular momentum, mass and energy of the halo at the shell
virialization time $t_v$. The spin parameter, $\lambda$, is assumed to 
be constant in time. As the result of the assembling of these 
mass shells, a present day halo ends with an angular momentum distribution 
close to the universal distribution found by 
Bullock et al. (2001). A fraction \fd\ of the mass of each shell is assumed 
to cool down and form a disk in a dynamical time. The radial mass distribution 
of the infalling gas is calculated by equating its specific angular 
momentum to that of its final circular orbit in centrifugal equilibrium 
(detailed angular momentum conservation is assumed). The gravitational 
interaction of disk and halo is calculated using the adiabatic invariant 
formalism. As the result of this modeling, disks with a nearly exponential 
surface density distribution are formed (Firmani \& Avila-Reese 2000, hereafter
FA00; Avila-Reese \& Firmani 2002). The local star formation (SF)
is triggered by the Toomre gas gravitational instability criterion and 
self-regulated by a vertical disk balance between the energy input due to SNe 
and the turbulent energy dissipation in the ISM. 
The SF efficiency depends on the gas surface 
density determined mainly by $\lambda$, and on the gas accretion 
rate determined by the cosmological MAH. 
Finally, we consider the formation of a secular bulge using the Toomre 
criterion for the stellar disk.

In this paper we use a flat model with non-vanishing vacuum 
energy density ($\Omega_{\Lambda}=h=0.65$) and the power spectrum 
normalized to COBE.

\section{Shapes of rotation curves}

Although the outer model RCs are not always flat, their shapes are smooth,
showing a conspiracy between the dark and luminous matter distributions. 
The maximum of the RC is typically attained at 2-3 stellar scale radii, R$_s$.
For a given mass, {\it the outer shape of the RC correlates with the 
disk SB, the bulge-to-disk (b/d) ratio, and the galaxy mass fraction 
$\fd\equiv$ M$_{\rm gal}/$M$_v$}.
The MAH does not play major role in the outer RC shape but, for a given mass, 
it defines the amplitude of the RC (\vmax) (see Fig. 4 in FA00).
 
$\bullet$ We find that the higher the disk SB is, the steeper the declining 
shape of the RC (Fig. 1) and the larger the b/d ratio. 
In Fig. 2(a) we plot the outer slope of the RC, $lgsl$, vs the stellar central 
surface density for 20 models of M$_v=3.5 \times 10^{11}\msun$, \fd=0.05, 
and the MAH and $\lambda$ calculated randomly from the corresponding statistical
distributions (see FA00). Filled and empty symbols are for
models with b/d ratios larger and smaller than 0.1, respectively. $lgsl$ is 
the logarithmic slope of the RCs at 2.2 and 5 R$_s$. 
Some observational works found similar correlations for the family
of high SB galaxies (e.g., Casertano \& van Gorkom 1990; Verheijen 1997, who 
also included low SB galaxies), and in the extreme of low SB, it is well
known that low SB galaxies have typically rising RCs (de Blok \& McGaugh
1997). As $\lambda$ is smaller, the gaseous disk is more 
concentrated and the stellar disk is more unstable giving rise to 
high SBs and large b/d ratios as well as to more peaked RCs. As seen in Fig. 1, 
for the range of the most probable
values of $\lambda$ and for $\fd\sim 0.05$ this may happen in spite of the 
CDM halos being cuspy. Nevertheless, for these halos, only the more concentrated 
disks may resemble the maximum disk solution and only very few models (low SB)
present rising RCs. We will return to this question below.

\begin{figure}
\vspace{5.4cm}
\includegraphics{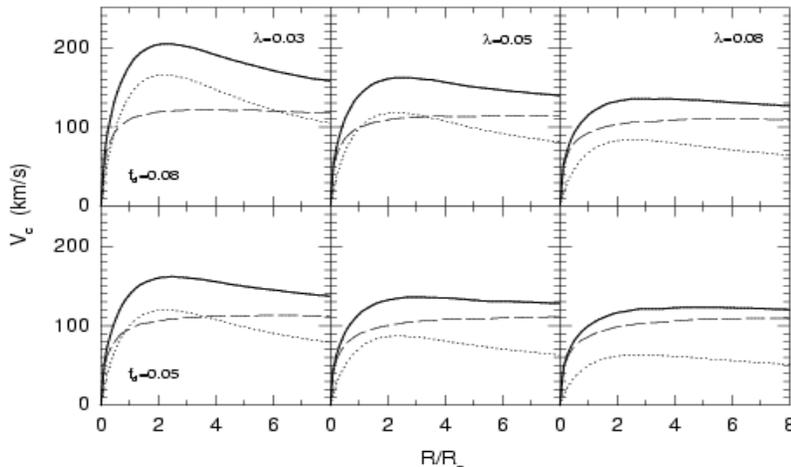}
\caption{Rotation curve decompositions for models of M$_v=3.5\times 10^{11}\msun$
and with the average MAH. The dotted and dashed lines are the disk and halo
components, respectively. Form left to rigth $\lambda$ increases (SB decreases)
and the upper and lower panels are for \fd=0.08 and 0.05, respectively. The
radii are scaled to the corresponding stellar scale radius R$_s$ ; \vmax\ occurs
typically at 2-3 R$_s$.}
 
\end{figure} 

$\bullet$ In Fig. 1 one sees that as \fd\ gets larger, the 
more declining is the outer RC slope. In our models the fraction of galaxy mass 
is a free parameter. The gas cooling efficiency
within the dark halos can be a function of mass; in large halos the 
cooling time tends to be larger than the Hubble time in such a way that 
only a fraction of the original baryons could be in the $z=0$ 
galaxy. At the other extreme, very small systems may lose significant 
fractions of gas due to SF activity (SNe). 
Calculations taking into account these two processes indeed show that 
\fd\ may be $\sim 0.5$ times the original baryon-to-dark matter fraction in 
the CDM halos (van den Bosch 2001). These calculations show that the final 
\fd\ of the halos does not follow a particular trend 
with M$_v$; if any, smaller halos have on average smaller 
\fd's than larger halos. On the other hand, the original baryon-to-dark
matter fraction in the halos can not be larger than the universal fraction, 
which for $h=0.7$ has a central value of $\Omega_b/\Omega_m\approx 0.15$. 
Cosmological simulations have 
recently shown that due to the gas heating in the shocks during the formation 
of filaments, only roughly one third of the baryons are trapped within 
collapsed galaxy halos at $z=0$ (Dav\'e et al. 2001). Therefore, on the 
ground of all these theoretical calculations, one may expect that the 
galaxy-to-halo mass fraction in halos is on average 
$\fd\approx 0.15/(2\times 3)=0.025$. Estimates of this fraction for our 
Galaxy and other galaxies indeed suggest similar values.

Modelers have commonly used values of 0.05 for \fd\ (e.g., Mo et al. 1998; 
FA00). For this value, model disks with $\lambda\lesssim 0.025$ 
($\approx 10\%$ of the $\lambda$ distribution) have typically highly declining 
RCs to be realistic. Since disk component dominates strongly in these cases, 
the disk is dynamically unstable and probably turns into a spherical 
system. As \fd\ gets smaller, the disk component becomes less
dominant (Fig. 1) and, therefore, models with smaller $\lambda$'s can
still be stable. On the other hand, for \fd=0.05 low SB galaxies will be 
those with $\lambda > 0.055$ ($\approx 45\%$ of the $\lambda$ distribution). 
The SB distribution of  disk galaxies shows that more than 
50\% of them are probably of low SB type (de Jong \& Lacey 2000).
Using $\fd\approx 0.03$, (i) the low SB galaxies are those with 
$\lambda \grtsim 0.035$ (now more than 50\%) and (ii) high SB models with  
$\lambda \approx 0.015$ have still realistic RCs.

\begin{figure}
\vspace{5.cm}
\includegraphics{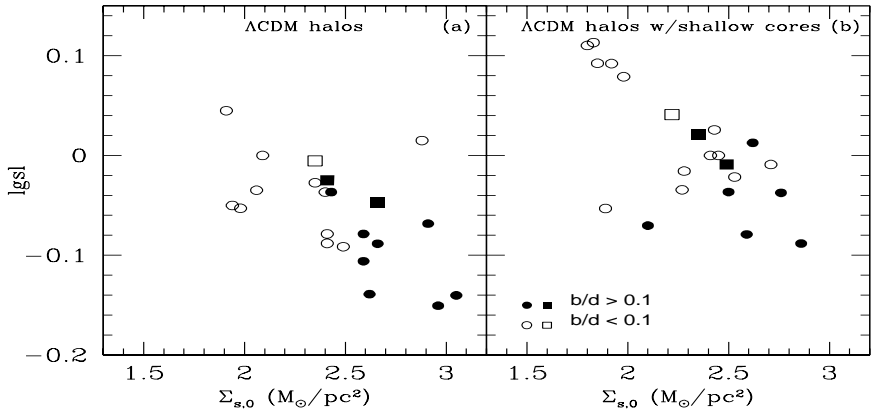}
\caption{Outer logarithmic slope of RC vs the stellar central surface density
of 20 models with the same M$_v$ and \fd, but with MAHs and $\lambda$ calculated
randomly for the corresponding statistical distributions (circles). Squares
are models with the same MAH, $\lambda$, and \fd, but 3 different masses, which
increase from left to rigth. }

\end{figure} 

$\bullet$ Finally, we find that the outer RC correlates slightly
with mass (luminosity) in the sense that massive systems tend to have 
declining RCs at one optical radius for example, while for the smallest 
systems the RCs are slightly rising at this radius. At first glance this could 
seem to be in conflict with the CDM halos: the less massive halos are more 
concentrated than the more massive ones. However, the central surface density of 
the model disks increases with the mass as M$_v^{a}$ with $a\approx 0.2-0.3$ 
(see also Dalcanton et al. 1997), and as discussed above,
the outer shape of the RCs correlates with the disk surface density. In Fig. 2 
squares are models with the average MAHs, $\lambda=\fd=0.05$, but with 3 different
masses, M$_v=3.5 \times (10^{10}, 10^{11}, 10^{12})\msun$, from left to right,
respectively. The plot shows that {\it the mass dependence of $lgsl$ is mainly due 
to the dependence of  M$_v$ on SB}. From
an analysis of a large sample of late-type galaxy RCs, Persic et al. (1996) 
concluded that RCs follow an universal trend which depends only
on luminosity, L. The RCs of our models could be more dependent on L
if \fd\ would depend on mass. We have mentioned above that for small 
galaxies (dwarfs), \fd\ could be on average smaller than for massive ones;
feedback is probably important for the small galaxies.

We conclude that model galaxies have different RC shapes, which correlate 
primarily with the disk SB (determined mainly by $\lambda$) and \fd. A 
dependence on L as suggested by Persic et al. (1996)
may appear but as a secondary effect, due to the dependence of the
SB and \fd\ on the mass. A more careful comparison of model and observed
RCs could allow us to put some constraints on the value and dependence
on mass of the unknown parameter \fd.

\subsection{Rotation curve decompositions}

From Fig. 1 one sees that the halo component dominates in the RC
decomposition of models with \fd=0.05 and $\lambda\grtsim 0.035$. 
In other words, models with SB typical for normal high SB galaxies are 
sub-maximal (FA00; Avila-Reese \& Firmani 2002). If $\fd\lesssim 0.05$, as 
theoretical and observational arguments hint (see above), then even for the 
highest SB galaxies the halo component will be dominant in 
the center, i.e. all the models are sub-maximum disks, the low SB
ones tending to the minimum disk case. Observations suggest
that HSB galaxies tend to the maximum disk case (e.g., Corsini et al. 
1998; Salucci \& Persic 1999; Palunas \& Williams 2000). Comparison of 
population synthesis models with the photometric properties of HSB 
galaxies also point to mass-to-luminosity ratios corresponding to 
the maximum disk case (e.g., Bell \& de Jong 2001). Theoretical arguments 
as the swing amplifier constraints also suggest central luminous matter 
dominion in HSB galaxies. 

The model RC decompositions are dominated by the dark component because 
the CDM halos are cuspy. Observations of the RCs of dark-matter-dominated
galaxies point to halo central mass distributions that are less concentrated 
than what CDM predicts. Recent high resolution H$\alpha$ (Blais-Ouellete et al. 2000;
see this volume), HI (C\^ot\`e et al. 2000) and CO (Bolatto et al. 2002) RC 
observations
for dwarf galaxies, and HI and H$\alpha$ RC observations for low SB galaxies
(de Blok et al. 2001; Marchesini et al. 2002) show that the inner density
profile of halos is much shallower than r$^{-1}$. Using data for dwarf 
and LSB galaxies and cluster of galaxies, Firmani et al. (2001) inferred
the scaling laws of the halo cores: the central density does not depend on 
mass, and the core radius increases roughly proportional to \vmax. Using 
these scaling laws, we artificially smoothed the central density profiles of 
our CDM halos. The RC decompositions obtained for these shallow CDM halos
show central dominance of the disk component over the halo for models 
with $\lambda\lesssim 0.05$ and 0.035 when $\fd =0.05$ and 0.03, respectively.
In other words, high SB model disks are now close to the maximum disk
solution (more the models with  $\fd =0.05$ than those with $\fd =0.03$).
As the SB decreases, the models are more and more dark-halo-dominated.
Low SB galaxies indeed seem to be sub-maximum disks; a maximum disk solution 
for these galaxies demands too high M/L ratios from the stellar 
population point of view (de Blok et al. 2001). Also, the outer logarithmic
slope of RCs of model galaxies increase on average when the soft core is 
introduced, in better agreement with observations (panel b in Fig. 2).

\section{Correlation among the residuals of the L-\vmax\ and L-R$_s$ 
relations: are disk galaxies sub-maximal?}

As mentioned in the introduction, if dark matter dominates strongly, 
then variations in the disk density will not affect significantly the
shape and amplitude of the RC, and in particular the \vmax, for a given 
disk mass. Courteau \& Rix (1998; hereafter CR98) studied this question 
in a statistical sense by using the Tully-Fisher (TF) and L-R$_s$ relations. 
They considered that if the halo component dominates for most of galaxies, 
then the residuals of the TF relation should not correlate with the 
residuals of the L-R$_s$ relation, i.e. with the SB of the disk. CR98 analyzed 
two large observational samples of high SB galaxies
and found a very small correlation among the residuals of these relations.
So they concluded that high SB galaxies are sub-maximum disks.
On the other hand, observations suggest a dependence of the shape
of RCs with SB (see \S 3). It would seem that these two observational 
results are at odds one with the other. 

According to the models, the shape of RCs may depend on the SB (Fig. 1) 
and, at the same time, the residuals among the infrared TF and L-R$_s$ relations 
may not be correlated. In Fig. 2(a) we show $\delta$lgV$_{m}$, the
residual of the TF relation vs. $\delta$lgR$_s$, the residual of the L-R$_s$
relation. The filled circles are for high SB models while empty
circles are for low SB models. The observational sample used by 
CR98 has values of $\delta$lgR between -0.3 and 0.3. Within
this range there is a good agreement between models and observations; 
high SB galaxies show a small negative gradient while for low SB models
this behavior is reversed. On average, for galaxies of typical
SBs (not very high or not very low SBs) the correlation vanishes. This is 
because, for a given mass,  not only \vmax\ (or V$_{2.2}$) decreases as the SB 
decreases, but also the stellar disk mass, M$_s$ (or L); according to our 
SF mechanism, the efficiency of transformation of gas into stars 
is lower as the surface density is smaller. In panel (b) of Fig. 2 we 
plot the non-disturbed galaxies from the homogeneous and complete sample of 
galaxies of the Ursa Major cluster from Verheijen (1997). The observations 
follow roughly the same behavior of the high and low SB models! 

The argument that dark halo dominance in the RC decomposition implies
no correlation among the residuals of the TF and L-R$_s$ relations (CR98) applies
for the total disk mass (baryonic mass), i.e. when instead of L
(or M$_s$) we use the total (stars+gas) disk mass,  M$_d$.
When one passes from  M$_d$ to L or M$_s$ the compensation effect due
to the dependence of SF efficiency on SB eliminates the correlation
among the residuals of the TF and L-R$s$ relations if it exists. This is why 
also the residuals of the TF relation are not correlated with SB, and the TF 
relation of high and low SB galaxies is roughly the same (see FA00). 
In panel (c) of Fig. 2 we plot the residuals of the M$_d$-\vmax\ 
and M$_d$-R$_s$ relations for the same models of panel (a) (where $\fd=0.05$
was used). The average slope 
of the correlation is $\approx -0.35$ (this slope is $-0.5$ for maximum 
disks). If a shallow core is introduced (see \S 3.1), then  the slope is 
steeper, $\approx -0.4$. For $\fd\approx 0.03$ the slope results close
to $-0.2$ and $-0.25$ for the cuspy and shallow CDM halos, respectively.

Do observations show a correlation among the residuals of the M$_d$-V$_{m}$ 
and M$_d$-R$_s$ relations? In order to answer this question we constructed
the baryonic TF and M$_d$-R$_s$ relations using the data from Verheijen (1997)
with a distance to Ursa Major of 20.7 Mpc.
The residuals from these relations are shown in Fig. 3(d). In order to calculate 
M$_d$ we used the fiducial M/L$_K$-($B-R$) relation presented in Bell \& de Jong 
(2001), and for the LSB galaxies we used the M/L$_K$ obtained by Verheijen (1997)
for the unconstrained fits of the RCs to an isothermal model. The gas mass
was estimated from the integrated HI flux, taking into account the fraction
of He and H$_2$. The approximate slope of the correlation among the residuals 
is $-0.25$. This is in agreement with our models with $\fd\approx 0.03$ and
shallow cores, although due to the large uncertainties in the observational derivation
we do not think that we are able to constrain between cuspy
or shallow halos. If \fd\ is much larger than 0.03, the observational inference
is pointing to dark-matter-dominated galaxies, i.e. cuspy halos. Nevertheless,
we have already seen that several arguments suggest $\fd\lesssim 0.03$.

\begin{figure}
\vspace{6.cm}
\includegraphics{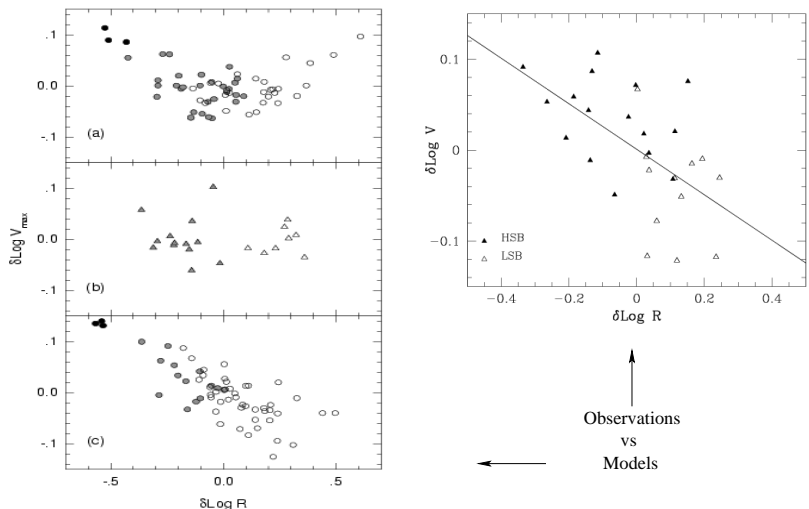} 
\caption{Residuals among the TF and L-R$_s$ relations for models (a) and
observations (b), and among the baryonic TF and M$_d$-R$_s$ relations for
models (c) and observations (rigth panel). The line is a linear regresion
to points (with slope -0.25). Solid and open symbols are HSB and LSB galaxies, 
respectively. See text for details.}
\end{figure}

\section{Conclusions}

$\bullet$ The outer RCs of disk galaxies formed within $\Lambda$CDM halos,
assuming detailed angular momentum conservation, tend to be more declining
as the SB, b/d ratio, galaxy mass fraction \fd, and mass (luminosity)
gets larger. The dominant correlation is the SB one and the key parameter
is $\lambda$. Since the mass correlates with SB, then the RC shape correlates 
with L, but an extra 
correlation of \fd\ with mass (feedback?) is necessary to reproduce
Persic et al. (1996) synthetic RCs. 
For \fd=0.05 the halo component dominates typically in the RC 
decomposition and only the highest SB models tend to be maximum
disks.

$\bullet$ Several arguments suggest $\fd\sim 0.03$, with
probably a small dependence on mass. For this
case, given the $\lambda$-distribution, more than 50\% of disk
galaxies are of low SB and the RCs are still realistic for $\lambda\approx 0.015$. 
On the other hand, even the highest SB models are sub-maximum disks. Introducing
shallow cores in the CDM halos, models with $\lambda\lesssim 0.035$ (high SB) are
close to the maximum disk solution. As the SB decreases, the model tends
to the minimum disk solution.

$\bullet$ Galaxy models of high and low surface density are segregated in the
baryonic TF relation. Therefore, the (anti) correlation among the residuals of 
the M$_d$-V$_{m}$ and M$_d$-R$_s$ relations is significant. However, when one
passes to the TF and L-R$_s$ relations the correlation almost disappears, but the 
correlation of the RC shape with SB remains, in agreement with observations.
This is because as the SB is higher, not only \vmax\ is larger for a given
mass, but the luminosity also (the SF efficiency depends 
on SB!). Thus, models of high and low SB shift along the main TF relation. 
The lack of correlation among the residuals of the L-\vmax\ and
L-R relations should not be interpreted as an evidence of sub-maximal disks, 
where the RC shape does not depend on SB. 

$\bullet$ A first comparison of the correlation among the residuals of the
baryonic TF and M$_d$-R$_s$ relations inferred from observations and models
suggest that if $\fd\grtsim 0.05$, then the observed galaxies are strongly 
dark-matter-dominated in the center. For the better preferred values of  
$\fd\lesssim 0.05$, the slope of the observed correlation is in better
agreement with models with shallow halos.

\acknowledgments V.A. thanks the organizers for financial support in order
to attend to the meeting, and J.C. Yustis for computational assistance. This 
work was supported by CONACyT grant 33776-E to V.A.

\end{document}